# Large Enhancement of Circular Dichroism Using an Embossed Chiral Metamaterial

S. Hamed Shams Mousavi [1], Sajanlal R. Panikkanvalappil [2], Mostafa A. El-Sayed [2], Ali A. Eftekhar [1], Ali Adibi [1]

1. School of Electrical and Computer Engineering, Georgia Institute of Technology, Atlanta, GA, 30332 USA

2. School of Chemistry and Biochemistry, Georgia Institute of Technology, Atlanta, GA, 30332 USA

**Abstract:**

*In the close vicinity of a chiral nanostructure, the circular dichroism of a biomolecule could be greatly enhanced, due to the interaction with the local superchiral fields. Modest enhancement of optical activity using a planar metamaterial, with some chiral properties, and achiral nanoparticles has been previously reported. A more substantial chirality enhancement can be achieved in the local filed of a chiral nanostructure with a three-dimensional arrangement. Using an embossed chiral nanostructure designed for chiroptical sensing, we measure the circular dichroism spectra of two biomolecules, Chlorophylls A and B, at the molecular level, using a simple polarization resolved reflection measurement. This experiment is the first realization of*

*the on-resonance surface-enhanced circular dichroism, achieved by matching the chiral resonances of a strongly chiral metamaterial with that of a chiral molecule, resulting in an unprecedentedly large differential CD spectrum from a monolayer of a chiral material.*

Optical activity is an intrinsic property of chiral molecules and molecular assemblies that lack mirror symmetry. It manifests itself as the rotation of linearly polarized light upon transmission or reflection, known as optical rotary dispersion (ORD). Equivalently, the difference in the response of the optically active molecules to two incident light beams with left-hand circular (LHC) and right-hand circular (RHC) polarization is known as circular dichroism (CD). Most biologically significant molecules, including proteins, sugars, and nucleic acids, possess optical activity [1-2]. CD spectroscopy provides quick and facile insight into the large-scale structure of the molecular systems due to the dependence of the CD spectrum on the three-dimensional molecular structure enables. In contrast, the information extracted from vibrational spectroscopy techniques, such as Raman and IR absorption, is usually more indicative of the material composition and the atomic bonds. As a result, CD spectroscopy has become a standard technique to interrogate the secondary and higher level structures of macromolecules and molecular systems in bio-related disciplines [3]. While large CD in most biomolecules is only observed at ultraviolet and low-to-mid visible wavelength ranges, some organometallic molecules, such as Chlorophylls and certain synthetic anti-cancer compounds, have large chiral response in the higher visible and infrared (IR) wavelengths [4-7].

CD is a typically weak process that stems from the interaction between the helicity of light (or spin angular momentum) and a chiral medium [8-11]. Due to the typically small chiral cross-section of the molecules, standard CD spectroscopy usually involves precise polarimetric measurements of a large number of target molecules in bulk or in high concentration solutions. Non-chiral plasmonic nanoparticles can modestly enhance CD of chiral molecules [12-15], simply on the premise of their local field enhancement. However, a greater enhancement of molecular CD can be achieved in the presence of an optical superchiral field (electromagnetic field with enhanced chirality), existing in the vicinity of a chiral metamaterial [16,17]. Chiral nanostructures preferentially interact with LHC or RHC light, depending on the wavelength and the nanostructure design, and can also enhance the optical activity of molecules within the local superchiral field [18-23]. Aside from sensing, photonic and plasmonic chiral nanostructures have been employed for polarization beam splitting [24-26], and generation of optical orbital angular momentum [9].

The chirality of an electromagnetic field can be quantified by a time-even pseudoscalar called optical chirality ($C$), which is a measure of the chiral asymmetry in the excitation of an optically active molecule [27,28] and is given by:

$$C = \frac{\varepsilon}{2} \vec{E}.\nabla \times \vec{E} + \frac{1}{2\mu} \vec{B}.\nabla \times \vec{B}, \quad (1)$$

where $\vec{E}$ and $\vec{B}$ denote the electric field and magnetic flux density, respectively, and $\varepsilon$ and $\mu$ are the permittivity and permeability of the environment. A circularly polarized plane-wave carries a non-zero optical angular momentum, therefore has a non-zero constant $C$. In the near-field of a resonant chiral metamaterial, however, the value of $C$ can be enhanced by several orders of magnitude. This property of chiral nanostructures can be brought into play in surface-enhanced circular dichroism (SECD) at the molecular level.

There are two types of chiral metamaterials, planar and three-dimensional chiral metamaterials. While planar chiral metamaterials can locally enhance chiral fields, they do not exhibit a CD spectrum in reflection or transmission measurements [29,30]. In strict terms, these metasurfaces are achiral, but they still possess some chiral properties in their near-field, which is sometimes called *planar chirality*. On the other hand, chiral metamaterials with 3D arrangement have typically stronger locally enhanced chiral fields and also have measurable CD spectra [20,21,23]. Previously, chiroptical sensing has been demonstrated using a planar chiral metamaterial [31]. The measured change in CD, in this experiment, was fairly modest due to the small chirality of the proposed planar metasurface, as well as the mismatch between the chiral resonance of the metamaterial and that of the target molecules. It has been suggested that in order to achieve a significant CD, at least a bilayer structure, with three-dimensional arrangement of plasmonic nanoantennas is needed [20,28]. However, these nanostructures usually are hard to fabricate at small scales and often need a dielectric cap layer to preserve the strong chiral response. Upon the removal of the cap layer, which is necessary for sensing applications, their chiral response is greatly reduced.

In this paper, we demonstrate a new three-dimensional chiral metamaterial that can be used for sensing applications without the limitations of the previous nanostructures. We show that the interaction between the metamaterial presented here and materials with significant chirality in the same range, can result in a large change in the chiroptical response of the overall system, beyond what could be achieved with a planar nanostructure. In addition this chiral metamaterial can be realized using a simple fabrication process with one-step lithography, that preserves the scalability of the embossed nanostructures necessary for practical applications at lower wavelengths.

In order to explain the origin of the chirality in our embossed nanostructure, we can use the Born-Kuhn (BK) model [32], in which, a bilayer nanostructure comprised of two closely spaced nanoantennas at an angle and in two vertical levels is presumed. The key feature of this nanostructure is the vertical separation between the two nanoantennas, which induces a phase difference between the two scattered waves and creates chirality. The strong coupling between the two nanoantennas generates two hybrid plasmonic modes, called bonding and anti-bonding modes, which interact differently with the two circular polarizations of light. However, the realization of this bilayer nanostructure would require stringent lateral alignment, similar to other multi-layer stacked metamaterials [33-35], albeit to a lesser degree since the structure has only two layers. This makes it difficult to scale the nanostructure to shorter wavelength range, where most molecules of interest show significant CD response. To solve this problem, we take advantage of the optical properties of a self-aligned vertical nanoantenna-nanowall-nanoaperture stack, to design a BK-type embossed chiral metamaterial with a one-step lithography process, developed previously for the fabrication of plasmonic nanostructured used for surface-enhanced Raman spectroscopy (SERS) [36].

In Fig. 1.a, we have shown a schematic view of one of the two possible enantiomers of the proposed embossed chiral metamaterial (Enant D). Figure 1.b shows the same enantiomer fabricated using our self-aligned fabrication technique. This nanostructure can be decomposed into two pieces, $A_1$ and $A_2$, each having an arc-shaped plasmonic nanoantenna stacked on a nanoaperture with similar dimensions via a dielectric nanowall. The nanoapertures also have localized surface plasmon polaritons (SPPs), akin the nanoantennas, as it can be inferred from the Babinet's principle [37-40]. The nanoantenna-nanoaperture system has two localized SPP modes, with the electric field mostly concentrated around the nanoantenna in one SPP mode and around the nanoaperture in the other (Figs 1.c,d). For simplicity, in the remaining part of this paper, we call these two modes, nanoantenna and nanoaperture modes, respectively. By matching the resonance of the nanoantenna mode in one nanoantenna-nanowall-nanoaperture stack ($A_1$) with the nanoaperture mode of the other stack ($A_2$), we can get the same effect as having two identical nanoantennas at two vertically separated layers, as presumed in the BK model. A more thorough explanation of the nanostructure design is presented in the *Supplementary Information*. The dual nanostructure, Enant L (not shown here), is merely the mirror image of the structure shown in Fig. 1.a versus any plane perpendicular to the x-y surface. Evidently, all the optical properties are also mirrored, e.g., the local field distribution of Enant L with LHC incident light is the mirror image of the local field distribution of Enant D in response to the RHC light. The fabricated embossed meta-atoms (Enant D) are shown in Fig. 1.b. Figures 1.c-e show the electric-field (E-field) distribution in the vicinity of $A_1$, $A_2$ and Enant D, respectively at their resonance wavelengths, in response to RHC light, over the cylindrical surface (S) shown in the inset of Fig. 1.a. As it can be seen in these figures, in $A_1$ the E-field is concentrated near the nanoantenna, whereas in $A_2$ the field is predominantly concentrated around the nanoaperture. This difference in the local distributions of the fields of $A_1$ and $A_2$ results in the large CD response of the overall nanostructure.

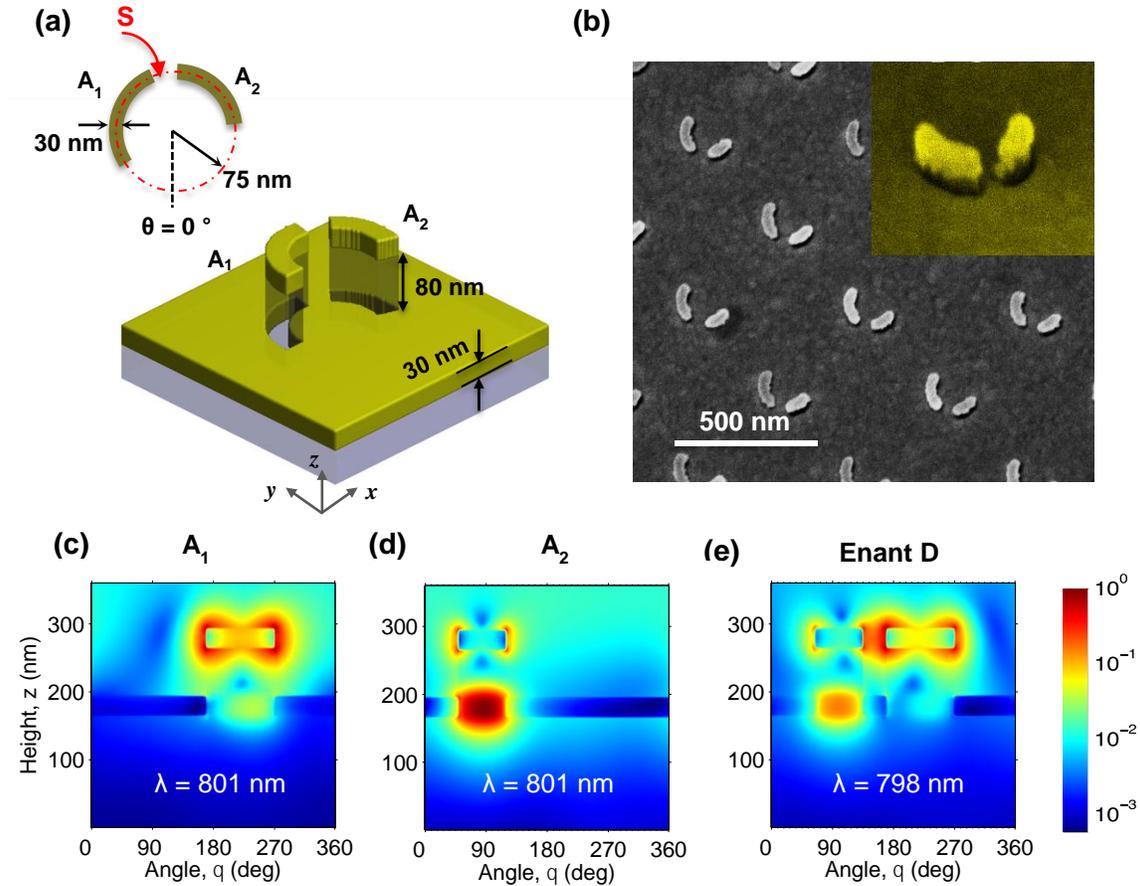

*Figure 1 Embossed chiral metamaterial. (a) Geometrical dimensions of a unit-cell of the embossed metamaterial (Enant D). The inner and outer radii of the two concentric arcs are 60 nm and 90 nm, respectively. The two arc-shaped nanoantenna-nanowall-nanoaperture stacks, $A_1$ and $A_2$, have the opening angles of 100° and 70°, respectively, and the separation angle between the two arcs is 35°. (b) Top-view scanning electron micrograph (SEM) of the embossed chiral nanostructure (Enant L). The inset is an oblique close-up view of Enant L at 45° tilt (false-colored). (c) E-field distribution of the nanoantenna mode in the larger nanoantenna-nanoaperture stack ($A_1$), in isolation, over the cylindrical surface (S) shown in the inset of Fig. 1.a at the wavelength of $\lambda$ = 801 nm (log-scale). (d) Similar E-field distribution for the nanoaperture mode of $A_2$. (e) E-field distribution for complete Enant D at $\lambda$ = 798 nm (resonance wavelength of Enant D in response to RHC light). The resonance wavelength of Enant D is*

*reduced to 798 nm due to the splitting (resonance wavelength of Enant L is at 810 nm, see the Supplementary Information, Fig. S.3.b).*

We have also calculated the chirality pseudoscalar *C* for our nanostructure to better visualize the local superchiral fields. In Figs 2.a,b, we have illustrated the distributions of the chirality enhancement factors, defined as $\Upsilon = C/C_{RHC}$ (with $C_{RHC}$ being the chirality of the plane-wave RHC light), in response to RHC and LHC light at λ = 781.5 nm at which the absolute value of the CD spectrum for Enant D is maximum as discussed later. The red and blue isosurfaces show the areas that have $\Upsilon = \pm 3.8$, showing the large difference in the response of the embossed nanostructure to RHC and LHC light. The local chirality for RHC light is largely enhanced around the larger nanoantenna and smaller nanoaperture, as it is expected from our design.

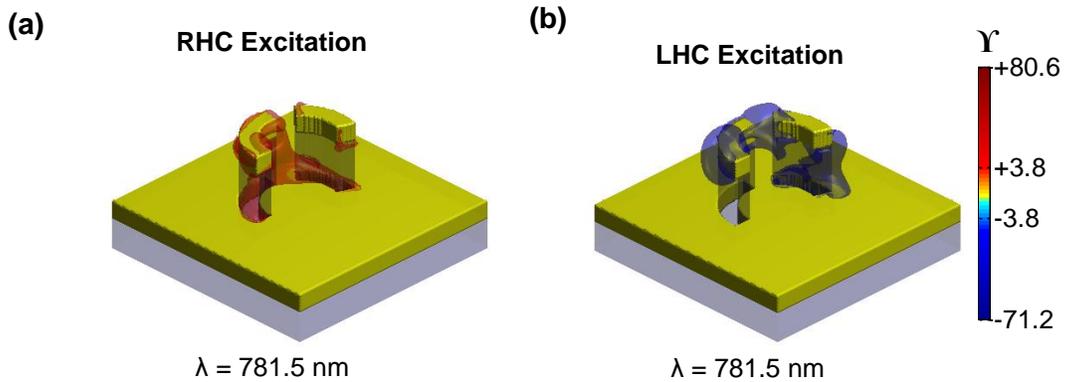

*Figure 2   Distribution of the chirality enhancement, $\Upsilon$, in the vicinity of Enant D, excited with (a) RHC and (b) LHC polarized light incident on the surface from the top. Blue and red surface shows the areas with $\Upsilon = \pm 3.8$ at λ = 781.5 nm, respectively (RHC resonance wavelength).*

To further investigate the potential of the proposed chiral nanostructures for sensing applications, two separate arrays of Enant D and Enant L are fabricated on a silicon substrate with a 6 μm-thick $SiO_2$ layer on top, and used for the spectral analysis of two chiral molecules, i.e. Chlorophyll (Chlor) A and Chlor B. The details of our fabrication process are explained in the *Supplementary Information*. In short, a thin layer of hydrogen silsesquioxane (HSQ) was spin-coated on the Si substrate and patterned using electron beam lithography (EBL) to form the arc-shaped dielectric walls. Consequently, both layers of the embossed metamaterials were formed simultaneously by depositing 2 nm of titanium, followed by the deposition of 30 nm of gold.

We used a polarization-resolved reflection measurement setup (Fig. 3) to measure the CD spectrum of the fabricated metamaterial before and after coating with the two types of Chlorophyll. Two sets of linear polarizers and quarter-wave-plates were used to generate RHC and LHC light beams in the excitation path and separate the two polarization components of the reflected light in the collection path. It should be noted that the circular polarization is reversed upon reflection.

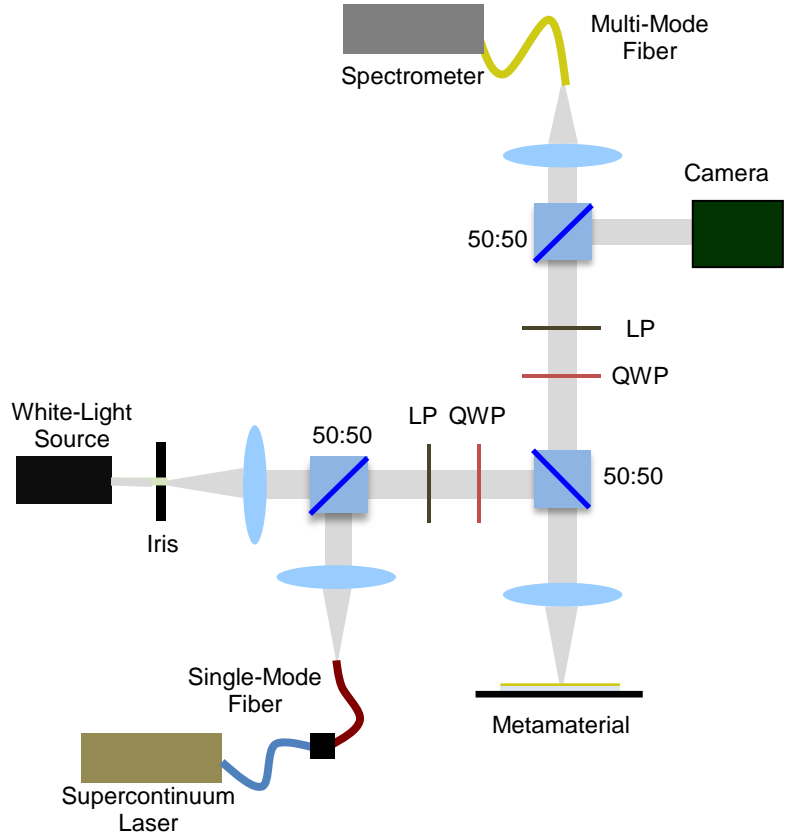

*Figure 3* Schematic diagram of the polarization-resolved reflection setup for the acquisition of the CD spectra of the metamaterials before and after coating. In this setup, the outputs of the white-light source and a supercontinuum laser are collimated and combined using a non-polarizing 50:50 beam-splitter (shown by 50:50). In the output, another 50:50 beam-splitter is used to direct two portions of the reflected light to the camera and the spectrometer. Two sets of linear polarizers (LPs) and quarter-wave-plates (QWPs) are used to generate RHC and LHC light beams in the excitation path and separate the two polarization components of the reflected light in the collection path.

The chiral response of the two enantiomers of the embossed nanostructure obtained from full-field numerical calculations (using finite-difference time-domain method) and experimental measurements are shown in Fig. 4. The blue and red curves in Figs. 4.a,b are the differential reflectance, i.e. $\Delta r = r_R - r_L$, for Enant D and Enant L, respectively, with $r_R$ and $r_L$ being the power reflection coefficients (ratio of the reflected power to the incident power) under RHC and LHC light. Both the simulation and experimental results in Fig. 4 show strong chirality that can be utilized for the CD-based spectral analysis of chiral biomolecules with high sensitivity. As it can be seen from Fig. 4.b, the differential reflectance measured from Enant L is not the exact opposite of the spectrum acquired from Enant D. This is due to the fabrication-induced variations in the sizes of the large and small arcs in Enant D and Enant L (*Supplementary Information*). Given the intricate design of our chiral nanostructure, very small changes in the geometrical parameters can induce the detuning of the nanoantenna or nanoaperture modes or can significantly change the coupling between the two adjacent localized SPPs. The measured CD spectrum also shows two anomalous dispersions with opposing polarities (i.e. change of $\Delta r$ from positive to negative and vice versa), corresponding to two interfering CD resonances at 761 nm and 801 nm. This type of anomalous dispersion, also known as cotton effect, is the characteristic change in the CD spectrum, in which CD or equivalently $\Delta r$ undergoes an abrupt change in the vicinity of an absorption band [41]. This effect has been observed in chiral molecules and its existence is predicted for chiral metamaterials, as well [42-44]. In the measured spectra shown in Fig. 4.b, the distance between these two cotton effects is enlarged due to fabrication-induced variations, and consequently $\Delta r$. Nevertheless, both enantiomers show a strong chirality and can be used for CD-based spectral analysis, independently.

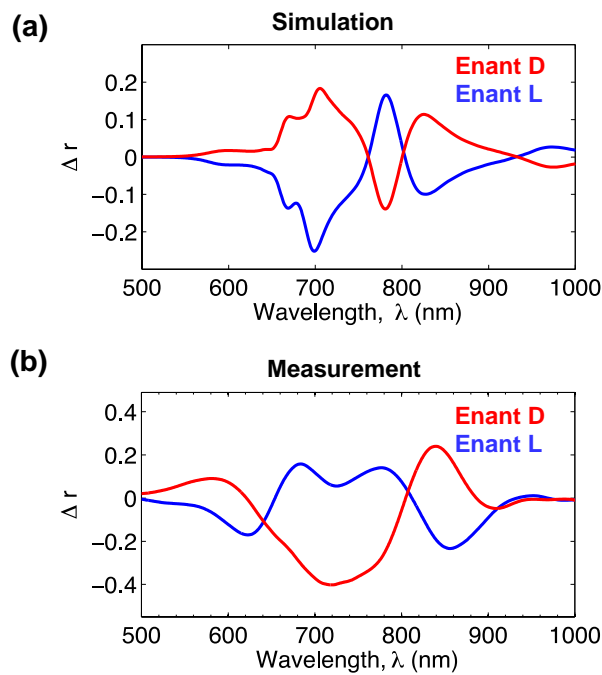

*Figure 4* *Differential reflectance, $\Delta r$, of Enant D and Enant L prior to coating with the chiral biomolecules from: (a) numerical simulations, and (b) measurements. All the geometrical parameters of the nanostructure are the same as those shown in Fig. 1.a.*

To perform chiroptical sensing experiments, we coated the surface of the two nanostructures by a monolayer of two naturally occurring organometallic molecules, Chlor A and Chlor B. These molecules are the main components of the light harvesting complexes that can be found in plants and micro-organisms and are responsible for capturing the sunlight and directing it to reaction centers to carry out photosynthesis [45]. Besides their strong absorption, Chlor A and Chlor B also exhibit large CD responses in visible and near IR wavlengths, with two CD resonance dips at $\lambda = 667$ nm and $\lambda = 652$ nm, respectively (*Supplementary Information, Fig. S.7.b*). These spectral features fall in the wavelength range that our embossed nanostructures also show strong chirality and the local superchiral fields can greatly enhance the optical activity of chiral biomolecules bound to their surface. We define the reflection mode CD as:

$$CD = 33(\log r_R - \log r_L), \quad (2)$$

in conformity with the definition of CD in standard transmission mode CD spectroscopy [2]. In addition, we define the differential CD spectrum ($\Delta CD$) as the difference between the CD spectra acquired from the coated and uncoated metamaterials ($CD_{MM+Mol}$ and $CD_{MM}$, respectively), i.e. $\Delta CD = CD_{MM+Mol} - CD_{MM}$. Figures 5.a-b show the measured change in CD spectra of Enant D and Enant L, when coated by Chlor A and B.

As it is evident in measurement $\Delta CD$ spectra in Fig. 5, both enantiomers show a giant change of several degrees in the CD by adsorption of the chiral molecules (i.e., Chlor A and Chlor B). This is a substantial improvement over the previously reported sensing devices that only show a CD change in the range of a few millidegrees [31]. This high sensitivity paves the way for precise CD-based spectral analysis for chiral molecules at very low concentrations. As it can be inferred from Fig.5, the measured $\Delta CD$ of the Chlor B-coated sample is stronger than the $\Delta CD$ acquired form the Chlor A-coated sample, which is because of the more pronounced CD of Chlor B in the near-IR wavelength range. Furthermore, we observe opposite behavior from Enant D and Enant L coated with Chlor A and Chlor B. We should also note that the $\Delta CD$ spectrum corresponding to Enant D (blue curves in Fig. 5) has a blue shift with respect to the spectra acquired from Enant L (red curves).

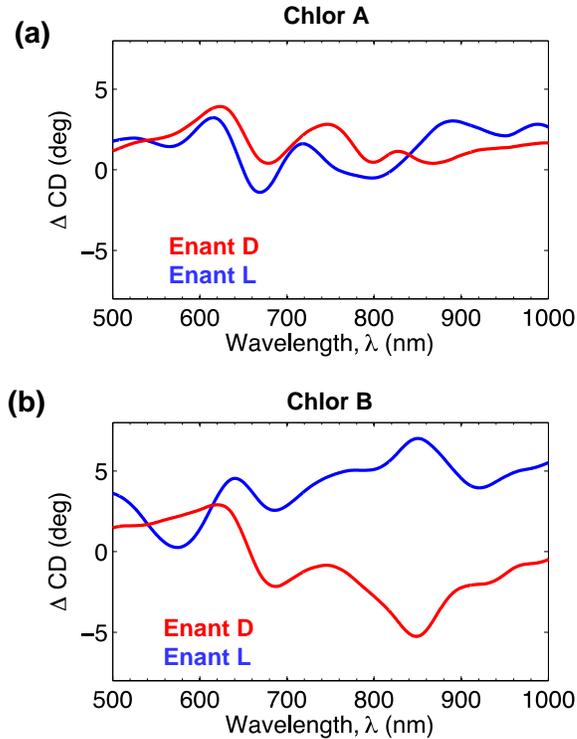

*Figure 5   Surface-enhanced circular dichroism (SECD) spectra acquired from the metamaterials coated with (a) Chlor A, and (b) Chlor B. The differential spectra, $\Delta CD$ is the difference between the CD spectra acquired from the coated and uncoated metamaterials. Blue and red curves correspond to Enants D and L, respectively. All structure parameters are the same as Fig 1.a.*

We have opted to use the complete $\Delta CD$ spectra for the differentiation of chiral molecules, since it provides a more quantitative measure for spectral analysis compared to only looking at the CD resonance shifts. In addition, the refractive index of the target molecules in this range has frequency dispersion and circular birefringence. Hence, modeling the change in CD, by a simple wavelength shift, $\Delta\Delta\lambda$, as it had been done previously [31], would not be very accurate.

In conclusion, we have demonstrated SECD-based spectral analysis of chiral biomolecules at the molecular level using an on-resonance 3D chiral metamaterial. Using these metamaterials, we measured the SECD signal from a monolayer of two biomolecules with very similar atomic compositions, and showed that we can distinguish them from the differential CD spectra acquired. The values of $\Delta CD$ reported in this work, show a very large improvement over the non-resonant experiments, performed using planar chiral metamaterial, nanoparticle assemblies or non-chiral nanoparticles, owing to the large chirality of the nanostructures and the spectral matching between the CD spectra of the metamaterials and target molecules. Similar experiments can be performed on other biomolecules that exhibit large CD response in the visible and near-IR wavelength range, including organometallic compounds with significant pharmaceutical applications (e.g. Ru-arene and Ferrocene-based anti-tumor drugs) [6,7], and rhodopsins with important applications in optogenetics [46-50]. Two stereoisomers of chiral compound can also be differentiated at the molecular level using the proposed nanostructure. Finally, given the scalability of this nanostructure, alternative metals such as silver and aluminum can be used to design metamaterials that have strong CD in low visible and even UV ranges of the spectrum, opening the door to the spectral analysis of natural supramolecules and molecular assemblies using SECD at the molecular level.

**Methods:**

**Numerical Simulations:**

Chiral metamaterials have been modeled using finite-difference time-domain (FDTD), to find the analytical differential reflectance, circular dichroism, and the chirality pseudoscalar. A thorough explanation of the numerical methods is presented in the *Supplementary Information*.

**Fabrication:**

The metamaterials were fabricated on a $SiO_2$ wafer (Si wafer with 6 μm of thermally grown $SiO_2$) . The oxide wafer was first covered by 110 nm of hydrogen-silsesquioxane (HSQ), which is a negative electron-beam resist and a spin-on dielectric material. Then, the thin HSQ layer was patterned using electron-beam lithography (EBL) to form the nanowalls, and 100 by 100 arrays with the lattice constant of 400 nm were formed on the oxide surface. Finally, 2 nm of titanium (Ti) and 30 nm of gold (Au) were deposited using electron-beam deposition to form the self-aligned bilayer nanostructure using a single lithography step.

**Sample Preparation:**

The target molecules, Chlor A and Chlor B, were coated on the nanostructure by immersing the sample in high concentration solutions. Two identical samples were soaked in 200 μM solution of Chlor A and Chlor B in pure methanol at room temperature and in the dark, for 10 minutes. At this concentration level, chlorophylls are believed to form a monolayer on the surface of the nanostructure [51]. Consequently, the sample is cleaned in pure menthol to remove unattached residual molecules.

The existence of self-assembled monolayer of chlorophylls was confirmed by performing surface-enhanced Raman spectroscopy (SERS) from the metamaterial using a 785 nm near-IR laser with approximately 2.64 mW incident power.

**Circular Dichroism Measurement of the Bulk Material:**

The CD spectra of Chlor A and Chlor B were measured using a commercial UV-visible spectropolarimeter (Jasco J-85) in the 300 nm to 800 nm wavelength range with 1 nm resolution. For bulk CD measurements, Chlor A was kept at the 200 μM concentration, but Chlor B was diluted to 50 μM due to its strong absorption. The measured spectra matched the results reported in ref [52].

**Measurement of the Surface-Enhanced Circular Dichroism:**

CD spectra of the metamaterials before and after coating were measured using a home-made polarization-resolved reflection measurement system. In our setup, RHC and LHC light were generated using a set of linear polarizer (LP) and quarter waveplate (QWP). A similar set of a LP and a QWP with reverse configuration was used at the output to separate the reflected RHC and LHC light. Other than the LPs and QWPs, all other components in the setup were approximately polarization and frequency independent to preserve the polarization content of light over a large spectrum.

*Author contributions:*

*S. H. Shams Mousavi, A. A. Eftekhar and A. Adibi conceived the idea. S. H. Shams Mousavi performed numerical simulations, fabricated the nanostructures, performed the surface coating, acquired and analyzed the SECD data. S. H. Shams Mousavi and Ali A. Eftekhar implemented the polarization-resolved reflection measurement setup. S. H. Shams Mousavi and S. R. Panikkanvalappil performed Raman measurements, and discussed the measurement approach with M. El-Sayed. S. H. Shams Mousavi performed the CD measurement of chlorophylls in bulk. S. H. Shams Mousavi, A. A. Eftekhar, and A. Adibi discussed the results at all steps. A. A. Eftekhar and A. Adibi guided the research. S. H. Shams Mousavi, A. A. Eftekhar and A. Adibi co-wrote the manuscript. All authors read the manuscript and provided comments.*

Properties of Chlorophylls in Oxygenic Photosynthesis—Succession of Co-Factors from Anoxygenic to Oxygenic Photosynthesis. by Z. Dubinsky, *Intech*. **3** (2013).

# Supplementary Information

## I. Design and Modeling of the Embossed Chiral Metamaterial

### I.1 Full-Wave Electromagnetic Simulations

Local chiral electromagnetic fields and circular dichroism (CD) spectra of the two enantiomers of the embossed metamaterial (Fig. S1.a,b) were calculated using 3D finite difference time-domain (FDTD) method (Lumerical Inc.). Two broadband plane-wave light beams with right-hand circular (RHC) and left-hand circular (LHC) polarization, propagating backwards in the z-direction, are used to excite the nanostructures from the top. The dispersive permittivity of gold (au) was modeled by fitting a sixth-order polynomial to the empirical data recorded by Johnson and Christy [1]. The simulation domain is confined by setting perfectly matched layer (PML) boundary conditions in the z-direction and periodic boundary conditions in the x and y. The mesh size is fixed at 1 nm in a cubic area with 400 nm edge length, centered at the surface of the substrate. Outside the high-resolution zone, a coarse adaptive mesh is used.

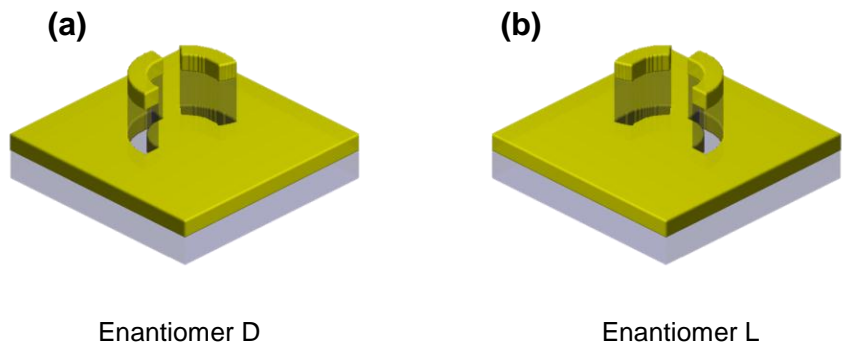

Enantiomer D        Enantiomer L

**Figure S.4   Schematic view of the embossed chiral metamaterials. (a) Enantiomer D, (b) Enantiomer L.**

The reflection-mode CD spectra are calculated by exciting the nanostructure with RHC and LHC light and recording the electric field (E-field) and magnetic field (H-field) of the backscattered light for both polarizations, using a frequency domain monitor placed above the source. Consequently, the RHC and LHC components of the reflection were separated to calculate the CD spectra. For instance, for an RHC excitation, we first calculate the E-field and H-field of the RHC component of the backscattered light from their $x$ and $y$ components using the Eq. S1,2.

$$\overline{E}_R = \sqrt{2}\left(E_x \hat{x} + iE_y \hat{y}\right) \qquad (S.1)$$

$$\overline{H}_R = -\sqrt{2}\left(H_x \hat{x} + iH_y \hat{y}\right) \qquad (S.2)$$

Then, the total reflected power, $P_R$, is calculated using the following relation:

$$P_R = real\left\{\int_S \left(\overline{E}_R \times \overline{H}_R^*\right)\hat{z}\,ds\right\} \qquad (S.3)$$

Assuming that the excitation source is purely right-handed, the reflectance for the RHC output component would be $r_R = \dfrac{P_R}{P_{exc}}$.

Similarly, by placing a LHC source, we can calculate $r_{LHC}$, using the total power of the LHC component of the reflected light. One measure of chirality, which can be evaluated experimentally for any given chiral system, is the differential reflectance, $\Delta r = r_R - r_L$, or equivalently the circular dichroism (CD) spectrum defined as:

$$CD_{ref} = 33\left(\log r_R - \log r_L\right) \qquad (S.4)$$

### I.2 Design of the Chiral Nanostructures

The embossed nanostructure can be decomposed into two pieces, shown in Fig. 2.a, b, each having an arc-shaped nanoantenna-nanowall-nanoaperture stack. Each vertical stack has two localized SPP modes, one with electric field mostly concentrated around the nanoantenna and the other around the nanoaperture. For simplicity, we call these two modes as the nanoantenna and nanoaperture modes.

In Fig. S.2.b, we have drawn the reflection spectra for two vertical stacks with the opening angles of 70 and 100 deg. The first and second resonance dips in each spectrum occur at the nanoantenna and nanoaperture resonances, respectively. By adjusting the opening angles, we are able to coincide the nanoantenna mode in one vertical stack, $A_1$, (first dip on the red curve) with the nanoaperture mode of the other stack, $A_2$, (second dip on the blue curve). Then, by placing these two arc-shaped stacks close to each other and adjusting the coupling between the two, we obtain a Burn-Kuhn (BK) type chiral nanostructure, shown in Figs S.1.a,b. In Fig. S.2.c- e, we have shown the distribution of the E-field at $\lambda$ = 801 nm in $A_1$, $A_2$ and at $\lambda$ = 798 nm in Enant D of the chiral nanostructure. The first wavelength, $\lambda$ = 801nm, is approximately at the first resonance dip of $A_1$, the second resonance dip of $A_2$ and and is close to the main resonance of Enant D, at $\lambda$ = 798 nm.

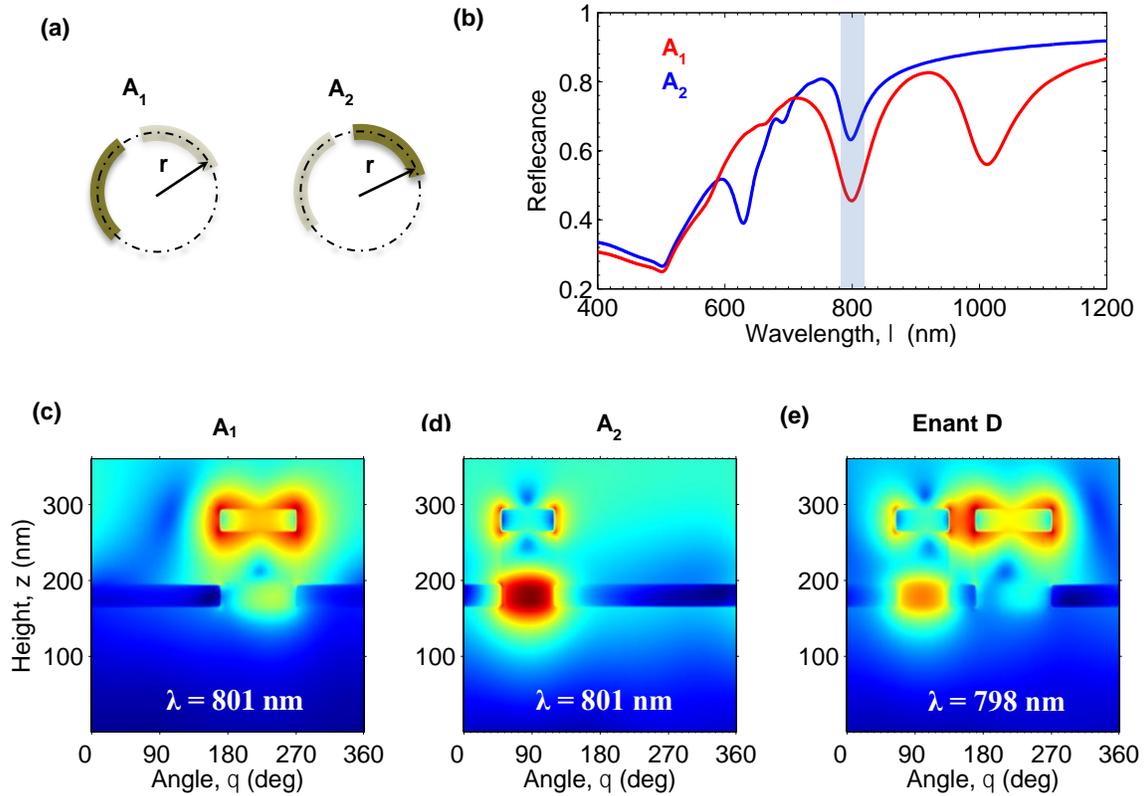

**Figure S.2** Design of the embossed chiral metamaterial, (a) schematic view of $A_1$ and $A_2$ in the complete Enant D structure (dark-colored arcs) from the top, (b) Reflection spectra of $A_1$ and $A_2$ in isolation; note the resonance mode matching between the nanoantenna mode of $A_1$ and the nanoaperture mode of $A_2$ at around $\lambda = 801$ nm, (c) E-field distribution of $A_1$ at $\lambda = 801$ nm showing the confinement of E-field around the nanoantenna, (d) E-field distribution of at $\lambda = 801$ nm showing the confinement of the E-field at the nanoaperture, (e) E-field distribution of Enant D at $\lambda = 789$ nm displaying the two virtual interacting layer, which result resulting in CD (according to BK model).

The separation angle between the two vertical stacks is a critical design parameter, as it determines the strength of the coupling between the two localized SPP modes. To find the optimal angle, the reflection for an enantiomer (Enant D) for the RHC and LHC source is numerically calculated for the range of the separation angle ($b$) from 5 to 95°. We found a design window at around $\beta = 35°$, for which the largest separation between resonance wavelengths, $\lambda_{res}$, for RHC and LHC polarization is observed (Fig S.3.a), corresponding to the maximum circular dichroism (CD).

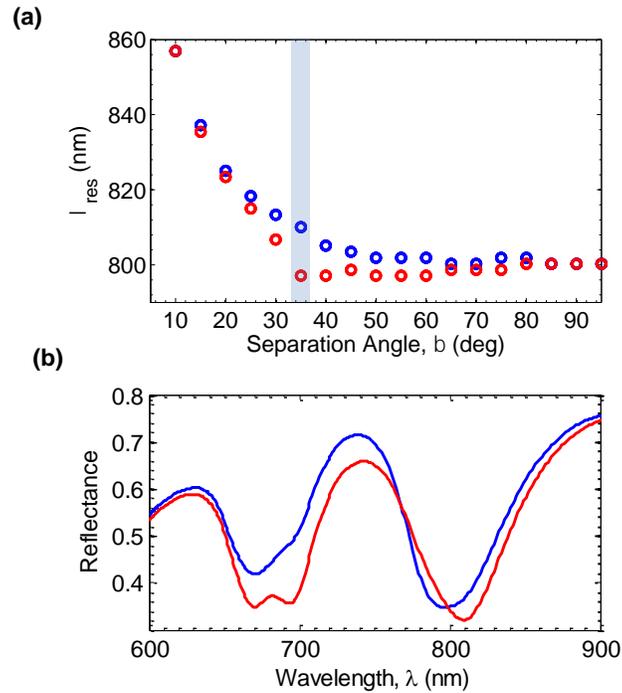

**Figure S.3 Optimization of the separation angle, $b$. (a) $\lambda_{res}$ of Enant D in response to RHC (blue circle) and LHC (red circle) light. The difference between $\lambda_{res}$ in two cases is maximized, when setting $b$ at 35°. (b) Resonance spectra of Enant D with the optimal $b$ for RHC and LHC excitation (blue and red curves, respectively).**

## II. Fabrication Process

Two separate arrays of enantiomers D and L were fabricated on a silicon substrate with a 6 μm-thick layer of oxide on top. The spacing between the meta-atoms in each 200 by 200 array was set at 400 nm, making the overall footprint of the arrays 80 μm by 80 μm. The thermal oxide was grown on a silicon wafer by wet oxidation in an oxidation furnace at 1100 °C. Consequently, the $SiO_2$ substrate was covered by a 110 nm-thick uniform layer of hydrogen-silsesquioxane (HSQ) by spin-coating XR-1541 resist (Dow Corning), containing a 6% solution of HSQ in methyl isobutyl ketone (MIBK), with the spinning speed of 6000 rpm for 60 seconds. HSQ is a spin-on dielectric material, used in semiconductor industry as an inter-level dielectric in multi-layer structures, and it is also a negative-tone electron beam resist that can be patterned by electron beam lithography (EBL) [2]. To pattern the HSQ layer, the sample was first prebaked at 90 °C for 3 min. Then, EBL was performed, using a JEOL JBX-9300FS EBL system, with the dosage of 3720 $\mu C/cm^2$. In order to resolve the patterns and form the nanowalls, the sample was immersed in a 25% aquatic solution of tetramethylammonium hydroxide (TMAH) for 30 seconds, and was consequently rinsed by deionized water for 2 minutes. The two layers of the nanostructure were finally formed by depositing 2 nm of titanium (Ti) as the adhesion layer and 30 nm of gold (Au) using electron-beam evaporation.

## III. Surface Adsorption of the Chlorophylls

Two identical samples were prepared for the coating of the two types of chlorophyll (Chlor A and Chlor B). First, we dissolved the two chlorophylls (Santa Cruz

Biotechnology, Inc.) in pure methanol at the fixed concentration of 200 μM, resulting in two solutions with distinct colors (Chlor A: green with blue tint, Chlor B: green with yellow tint) in agreement with the reports in the literature [3]. The surface coating was accomplished by immersing the chiral sample in the two solutions for 10 min in dark. Under these conditions, Chlorophyll is believed to form a monolayer on the surface of the nanostructure due to surface adsorption [4]. Then, the two samples were taken out of their respective solutions and were washed in pure methanol to remove any unabsorbed residual and dried with nitrogen gun.

The two solutions were characterized by performing absorption and CD spectroscopy (Fig. S.7). In the case of Chlor B, the solution was diluted to 50 μM, because of its strong absorption, for these measurements.

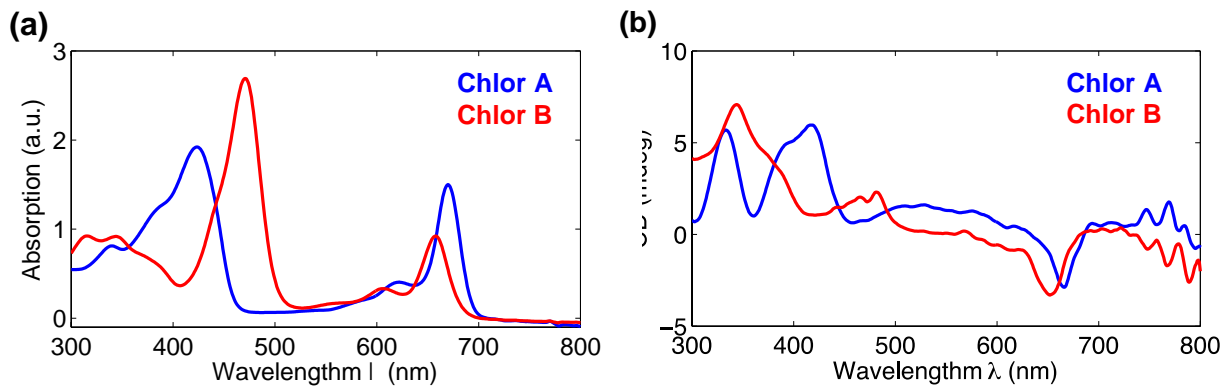

**Figure S.7    Characterization of the target molecules, (a) Absorption spectra of Chlor a (blue curve) and Chlor b (red curve), at the concentration of 200 μM and 50 50 μM, respectively, (b) CD spectra of Chlor a (blue curve) and Chlor b (red curve), at the same concentration levels.**

## IV. Molecular CD Measurements

The differential reflectance and the CD spectra of the chiral nanostructures before and after coating were measured using a polarization-resolved reflection measurement system .In this setup, we use a white light source for imaging. However, for CD measurements, a supercontinuum laser (SCL) is used. The output of the SCL is coupled to an optical fiber that is single mode over the target wavelength range to eliminate the higher order spatial modes of SCL, which can result in the spatial variation of polarization in the excitation region. In the excitation path, the RHC and LHC light beams are generated using a linear polarizer (LP) and a quarter-wave plate (QWP) with ± 45° difference between the polarization angle of LP output and the fast-axis of the QWP. Similarly, in the collection path, a QWP and a LP are used to separate the RHC and LHC components of the reflected light.  It should be noted that the circular polarization is reversed upon reflection. The measured CD spectra from each enantimer of the chiral nanostructure before ad after coating is presented in Fig. S.9.a,b.

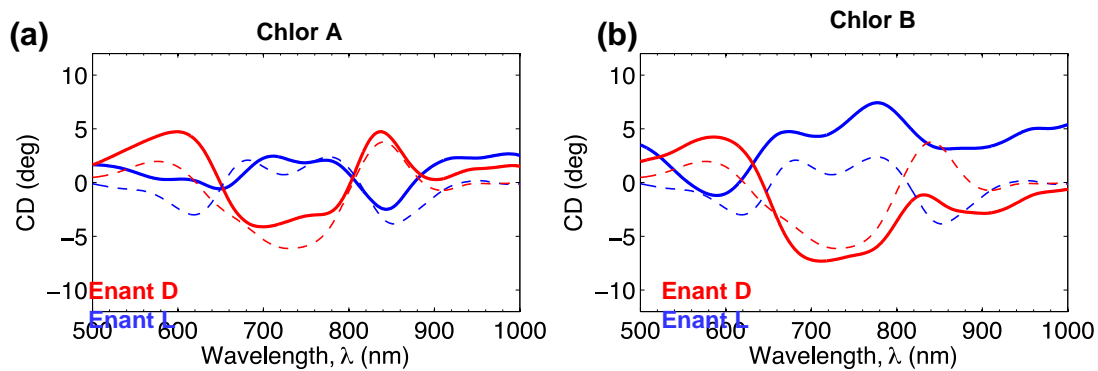

*Figure S.9    Measurement of surface-enhanced circular dichroism (SECD). (a) Schematic diagram of polarization-resolved reflection measurement setup for the measurement of the circular dichroism (CD) of the metamaterials before and after coating.  Surface-enhanced circular dichroism (SECD) spectra acquired from the embossed chiral nanostructures, (**b**) SECD spectra from Enant D (solid blue curve) and Enant L (solid red curve) coated with Chlor A; the blue and red broken lines show the CD spectra of Enant D and L, respectively, prior to coating with Chlor A, (**c**) Similar SECD spectra acquired from Enant D and Enant L before and after coating with Chlor B.*

# Supplementary Information References